\documentclass[12pt,twoside,british,sort&compress]{fec}
\usepackage[T1]{fontenc}
\usepackage[latin9]{inputenc}
\usepackage[a4paper]{geometry}
\usepackage{verbatim}
\usepackage{amsbsy}
\usepackage{graphicx}
\usepackage{esint}
\usepackage[numbers]{natbib}

\makeatletter
\author[1]{J.T. Omotani}
\author[1,2]{S.L. Newton}
\author[1]{I. Pusztai}
\author[1]{T. F\"ul\"op}
\author[3,4]{E. Viezzer}
\author[ ]{the ASDEX Upgrade Team}
\affil[1]{Department of Physics, Chalmers University of Technology, 41296 Gothenburg, Sweden}
\affil[2]{CCFE, Culham Science Centre, Abingdon, Oxon, OX14 3DB, UK}
\affil[3]{Max-Planck-Institute for Plasma Physics, Boltzmannstr. 2, 85748 Garching, Germany}
\affil[4]{Department of Atomic, Molecular, and Nuclear Physics, University of Seville, Avda. Reina Mercedes, 41012 Seville, Spain}

\doi

\emailID{omotani@chalmers.se}

\hypersetup{colorlinks=True,allcolors=blue}

\makeatother

\usepackage{babel}
\begin{document}
\global\long\def\nn{n_{\mathrm{n}}}
\global\long\def\zh{\hat{\boldsymbol{\zeta}}}
\global\long\def\pin{\boldsymbol{\Pi}_{\mathrm{n}}}
\global\long\def\fn{f_{\mathrm{n}}}
\global\long\def\tx{\tau_{\mathrm{CX}}}
\global\long\def\ffi{f_{\mathrm{i}}}
\global\long\def\nni{n_{\mathrm{i}}}
\global\long\def\rp{\rho_{\mathrm{pol}}}
\global\long\def\mi{m_{\mathrm{i}}}
\global\long\def\bp{B_{\mathrm{p}}}
\global\long\def\ti{T_{\mathrm{i}}}
\global\long\def\cdotnarrow{{\cdot}}
%\widowpenalty10000
%\clubpenalty10000

\title{Edge Momentum Transport by Neutrals: an Interpretive Numerical Framework}
\maketitle
\begin{abstract}

Due to their high cross-field mobility, neutrals can contribute to
momentum transport even at the low relative densities found inside
the separatrix and they can generate intrinsic rotation. We use a
charge-exchange dominated solution to the neutral kinetic equation,
coupled to neoclassical ions, to evaluate the momentum transport due
to neutrals. Numerical solutions to the drift-kinetic equation allow
us to cover the full range of collisionality, including the intermediate
levels typical of the tokamak edge. In the edge there are several
processes likely to contribute to momentum transport in addition to
neutrals. Therefore, we present here an interpretive framework that
can evaluate the momentum transport through neutrals based on radial
plasma profiles. We demonstrate its application by analysing the neutral
angular momentum flux for an L-mode discharge in the ASDEX Upgrade
tokamak. The magnitudes of the angular momentum fluxes we find here
due to neutrals of up to $1{-}2\;\mathrm{N\, m}$ are comparable to
the net torque on the plasma from neutral beam injection, indicating
the importance of neutrals for rotation in the edge.
\end{abstract}

\section{Introduction}

Momentum transport in the tokamak edge is a crucial issue since rotation
suppresses magnetohydrodynamic instabilities such as resistive wall
modes \citep{Hender2007} and flow shear in the edge suppresses turbulence,
leading to high-confinement mode (H-mode) operation \citep{Terry00}.
Neutral particles are always present in the edge of the confined plasma
volume, and despite their low relative density can contribute strongly
to transport due to their high cross-field mobility, as has been demonstrated
theoretically \citep{Hazeltine1992,Catto1994,Helander1994,Catto98,Fulop98_1,Fulop98_2,Fulop01,Fulop02,Helander2003,Omotani2016}.
The influence of neutrals on confinement has also been observed experimentally
\citep{Versloot2011,Joffrin14_JET,Tamain2015} and the poloidal location
of gas-fuelling, inboard versus outboard, has been seen to be important
for the low- to high-confinement transition threshold \citep{Fukuda00,Valovic02_COMPASS,Field2004}.

It has been shown previously that when neutrals dominate the angular
momentum transport, the radial electric field and hence toroidal rotation
can be self-consistently calculated \citep{Fulop02,Helander2003,Omotani2016}.
It was also shown that the neutrals can generate intrinsic rotation,
since a toroidal heat flux in the ions drives a radial flux of angular
momentum through neutrals, as explained in \citep{Helander2003}.
However, there will be other contributions, potentially of similar
or greater magnitude, for instance from turbulence, ion orbit losses
\citep{Shaing1989,Stoltzfus-Dueck12_PoP}, non-axisymmetric magnetic
fields \citep{Nave2010} or finite orbit width effects \citep{pusztai2016}.
It is therefore important to evaluate the momentum transport due to
neutrals from experimental profiles, so that the magnitudes of the
various effects can be compared. Analytical solutions \citep{Fulop02,Helander2003}
rely on asymptotic ordering of the collisionality into the Pfirsch-Schl\"uter
(high collisionality) or banana (low collisionality) regimes. However,
the conditions typical in the edge plasma produce order unity collisionality,
for example $\nu_{\mathrm{ii}}L_{\|}/v_{T}\sim0.2{-}0.5$ for the
profiles we show in Section \ref{sec:Estimating-size-of}, where $v_{T}$
is the thermal velocity, $\nu_{\mathrm{ii}}$ the ion-ion collision
rate and $L_{\|}=\oint d\theta\,\left(dl_{\|}/d\theta\right)/2\pi$
the parallel connection length. In order to relax this restriction
numerical solutions of the drift-kinetic equation must be used. Recently,
the radial electric field and toroidal rotation have been calculated
numerically from the constraint that the angular momentum flux through
neutrals vanishes in steady state in the absence of external torque
\citep{Omotani2016} that is, as noted above, assuming that the neutrals
dominate the momentum transport and also that the radial gradient
of the neutrals is the dominant profile gradient. The latter assumption
makes the calculation of the momentum flux local (otherwise it would
depend on, for instance, the radial gradient of the toroidal rotation
and hence on second derivatives of the density, temperature, etc.).
Here, in order to avoid these assumptions, we introduce a new approach
which takes the background plasma profiles, including the radial electric
field, as given, allowing us to calculate the angular momentum flux
through neutrals directly, without requiring that they are the only
channel for momentum transport. This provides an interpretive tool
that can be applied to experimental data in order to compare the magnitude
of the neutral momentum transport to other mechanisms and to external
sources of momentum such as neutral beam injection (NBI) heating.
The interpretive approach is derived in Section \ref{sec:Neutral-momentum-transport}
and applied to an L-mode discharge from ASDEX Upgrade (AUG) in Section
\ref{sec:Estimating-size-of}. We discuss the implications of the
results in Section \ref{sec:Discussion}.

\section{Neutral momentum transport\label{sec:Neutral-momentum-transport}}

We outline here the calculation of the angular momentum flux through
neutrals, which are coupled by charge exchange to kinetic ions. Integrated
over a flux surface, the radial flux of toroidal angular momentum
carried by the neutral population is $V'\left\langle R\zh\cdot\pin\cdot\nabla\psi\right\rangle $
where $\pin=\mi\int d^{3}v\,\boldsymbol{v}\boldsymbol{v}\fn$ is the
stress tensor of the neutrals, $\fn$ is the distribution function
of the neutrals, $2\pi\psi$ is the poloidal flux, $V$ is the volume
enclosed by a flux surface, a prime denotes a derivative with respect
to $\psi$, $R$ is the major radius, $\zh=\nabla\zeta/\left|\nabla\zeta\right|$
with $\zeta$ the toroidal angle (increasing in the co-current direction),
and $\mi$ is the mass of the ions or neutrals. 

Our solution to the kinetic equation for $\fn$ relies on several
approximations. We take a short charge-exchange (CX) mean-free-path
(MFP) expansion; although the MFP is not always short compared to
profile length scales in the edge, it was found in \citep{Fulop01}
that the approximation is surprisingly accurate, compared to a full
solution allowing arbitrary MFP for a special class of self-similar
profiles. We neglect ionization, but this is most important for the
determination of the neutral density profile, which we take as an
input; ionization affects the momentum transport only by changing
the effective collision rate \citep{Catto98}. We use a simplified
CX collision operator \citep{Catto1994} that allows us to close the
neutral kinetic equation in an efficient way \citep{Catto98}, as
described below. These approximations allow us to avoid the computational
expense of Monte Carlo neutral codes, e.g. EIRENE \citep{Reiter2005eirene},
while retaining coupling to the kinetic ion distribution, rather than
only a drifting Maxwellian as implemented in \citep{Reiter2005eirene}.

The steady state neutral kinetic equation then takes the form~\citep{Catto98}
\begin{equation}
\boldsymbol{v}\cdot\nabla\fn=\frac{1}{\tx}\left(\frac{\nn}{\nni}\ffi-\fn\right),
\end{equation}
where $\nn$ and $\nni$ are the neutral and ion densities and $\ffi$
is the ion distribution function. $\tx^{-1}=\nni\left\langle \sigma v\right\rangle _{\mathrm{CX}}$
is the characteristic rate for charge-exchange interactions with $\left\langle \sigma v\right\rangle _{\mathrm{CX}}$
the thermal charge exchange rate,  where $\left\langle \sigma v\right\rangle _{\mathrm{CX}}=4.21\times10^{-14}\mathrm{\; m^{3}\, s^{-1}}$
for Deuterium ions and neutrals at $300\;\mathrm{eV}$ \citep{ADAS},
which is a typical temperature value of the profiles we use in Section
\ref{sec:Estimating-size-of}. Solving perturbatively for small $\tx v_{T}/L$,
where $v_{T}$ is the thermal speed and $L$ is a characteristic length
scale of the background profiles,
\begin{align}
\fn^{(0)} & =\frac{\nn}{\nni}\ffi,\\
\fn^{(1)} & =-\tx\boldsymbol{v}\cdot\nabla\left(\frac{\nn}{\nni}\ffi\right).
\end{align}
$\fn^{(0)}$ contributes a term to the angular momentum flux which,
being proportional to $\boldsymbol{\Pi}_{\mathrm{i}}$, is negligible
at $\mathcal{O}(\delta)$ in the gyroradius expansion \citep{Helander_book},
where $\delta=\rho/L$ and $\rho$ is the gyroradius. Thus we need
keep only $\fn^{(1)}$ and so
\begin{align}
V'\left\langle R\zh\cdot\pin\cdot\nabla\psi\right\rangle  & =-\mi\tx V'\left\langle R\int d^{3}v\,\left(\zh\cdot\boldsymbol{v}\right)\left(\nabla\psi\cdot\boldsymbol{v}\right)\boldsymbol{v}\cdot\nabla\left(\frac{\nn}{\nni}\ffi\right)\right\rangle \nonumber \\
 & \approx-\mi\tx\frac{d}{d\psi}\left(V'\left\langle \frac{R\nn}{\nni}\int d^{3}v\left(\zh\cdot\boldsymbol{v}\right)\left(\nabla\psi\cdot\boldsymbol{v}\right)^{2}\ffi\right\rangle \right),\label{eq:mtm-flux}
\end{align}
neglecting $\nabla^{2}\psi$ and using the identities $\left\langle \nabla\cdot\boldsymbol{A}\right\rangle =\frac{1}{V'}\frac{d}{d\psi}\left(V'\left\langle \nabla\psi\cdot\boldsymbol{A}\right\rangle \right)$
for any $\boldsymbol{A}$ \citep{Helander_book} and $\boldsymbol{v}\boldsymbol{v}:\nabla\left(R\zh\right)=0$%
\footnote{This identity follows from the fact that $\boldsymbol{v}\boldsymbol{v}$
is a symmetric tensor, while $\nabla\left(R\zh\right)$ is antisymmetric.%
}. Including the gyroradius correction, the ion distribution function
at the particle position $\boldsymbol{r}$ is
\begin{align}
\ffi(\boldsymbol{r}) & =f_{\mathrm{i,gc0}}(\boldsymbol{r})-\frac{e\Phi_{1}}{T_{\mathrm{i}}}f_{\mathrm{i,gc0}}(\boldsymbol{r})-\boldsymbol{\rho}\cdot\nabla f_{\mathrm{i,gc0}}(\boldsymbol{r})+g_{\mathrm{i}}(\boldsymbol{r}),
\end{align}
where $f_{\mathrm{i,gc0}}$ is a Maxwellian, $\boldsymbol{\rho}=\boldsymbol{r}-\boldsymbol{R}_{\mathrm{gc}}$
is the gyroradius vector with $\boldsymbol{R}_{\mathrm{gc}}$ the
guiding centre position, $\Phi_{1}=\Phi-\left\langle \Phi\right\rangle $
is the poloidally varying part of the electrostatic potential $\Phi$
and 0, 1 subscripts refer to the order in $\delta$. The first two
terms do not contribute to (\ref{eq:mtm-flux}) as they are isotropic
in $\boldsymbol{v}$. We obtain the non-adiabatic piece of the perturbed
distribution function, $g_{\mathrm{i}}=f_{\mathrm{i},gc1}+e_{\mathrm{i}}\Phi_{1}f_{\mathrm{i,gc0}}/T_{\mathrm{i}}$
from numerical solutions of the first order drift kinetic equation
using the \textsc{perfect} neoclassical solver (run here in radially-local
mode) \citep{Landreman14_PERFECT}, which assume that the flow is
subsonic, $\boldsymbol{V}\sim\mathcal{O}(\delta v_{T})$.

As noted in the introduction, this system was used to explore the
effects of magnetic geometry and collisionality on intrinsic rotation
driven by momentum transport through neutrals \citep{Omotani2016,Omotani2016b}.
In the next section we introduce the interpretive framework which
can be used to diagnose experimental results.

\section{Interpretive modelling of momentum flux\label{sec:Estimating-size-of}}

\begin{figure}[t]
\includegraphics[width=0.5\textwidth]{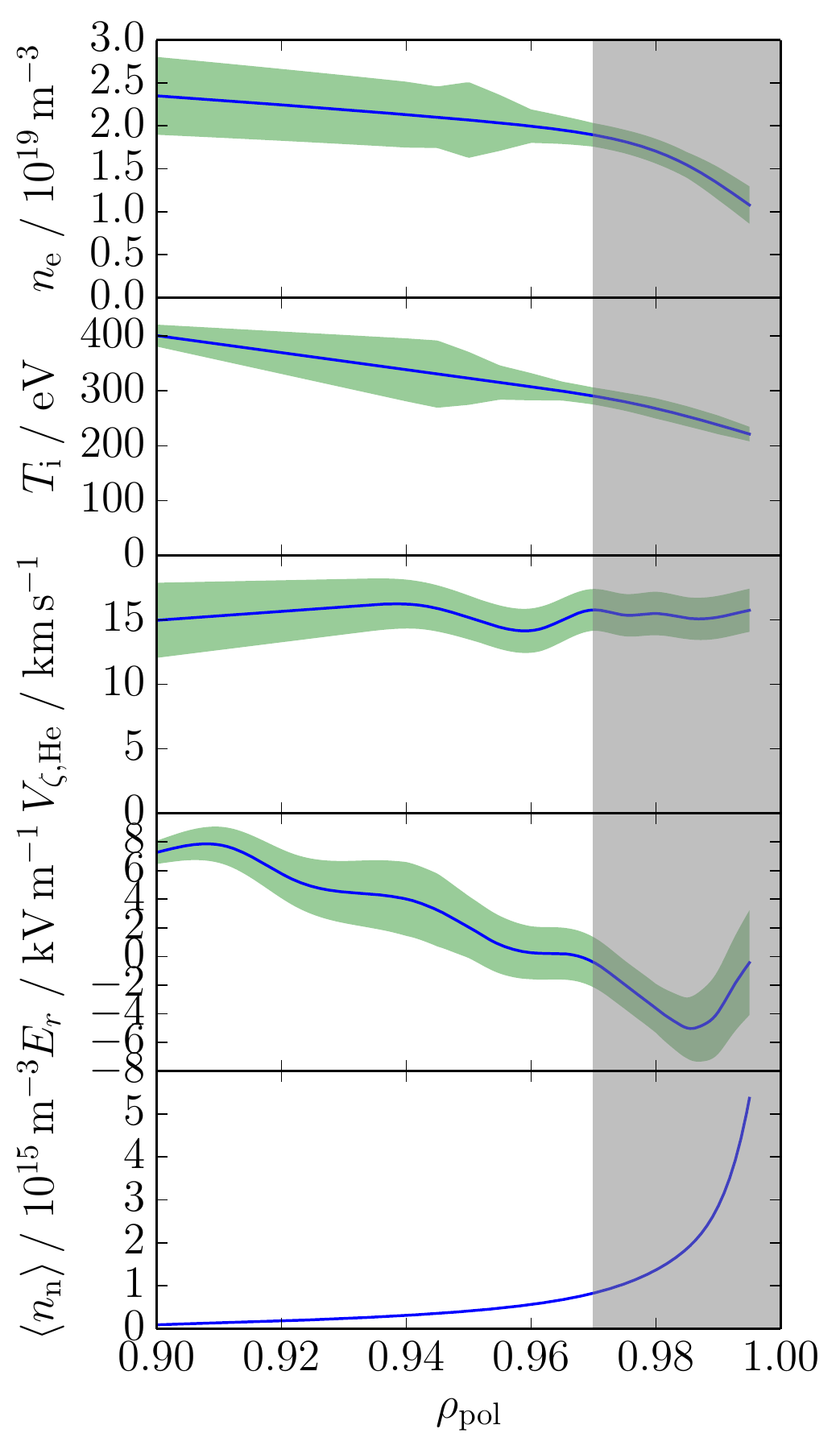}

\caption{Profiles from AUG discharge \#26601 as a function of $\protect\rp$.
From the top: ion density $\protect\nni$; ion temperature $\protect\ti$;
toroidal velocity of helium impurity $V_{\zeta,\mathrm{He}}$; and
radial electric field $E_{r}$ at the outboard midplane. Experimental
uncertainties are indicated by the green shaded regions. At the bottom
is the input profile we use for the flux surface average of the neutral
density $\left\langle \protect\nn\right\rangle $, from a KN1D simulation.
Grey shaded area is within a poloidal gyroradius of the separatrix.\label{fig:profiles}}
\end{figure}

We have evaluated the angular momentum flux through neutrals for L-mode
profiles from the AUG discharge \#26601, shown in FIG. \ref{fig:profiles},
in order to demonstrate the potential of our interpretive tool. This
takes profiles of ion density $\nni$ and temperature $\ti$, radial
electric field $E_{r}$ and neutral density $\nn$ as inputs to calculate
the angular momentum flux as described in the previous section. It
is also possible to constrain the neoclassical solutions with the
toroidal rotation profile of any ion species instead of $E_{r}$.

For the AUG case, CXRS measurements were performed on the $\mathrm{He}^{2+}$
impurity to measure the ion temperature (assumed equal to the $\mathrm{He}^{2+}$
temperature) and infer the radial electric field \citep{Viezzer2013}.
Electron density was measured with the Thomson scattering, Lithium
beam and interferometry diagnostics; we neglect the impurity density,
taking $\nni=n_{\mathrm{e}}$. Profiles are plotted as functions of
the normalized flux label $\rho_{\mathrm{pol}}=\sqrt{\psi/\psi_{\mathrm{sep}}}$
where $\psi_{\mathrm{sep}}$ is the value of $\psi$ at the separatrix
and we take $\psi=0$ at the magnetic axis.

\begin{figure}[t]
\includegraphics[width=0.5\textwidth]{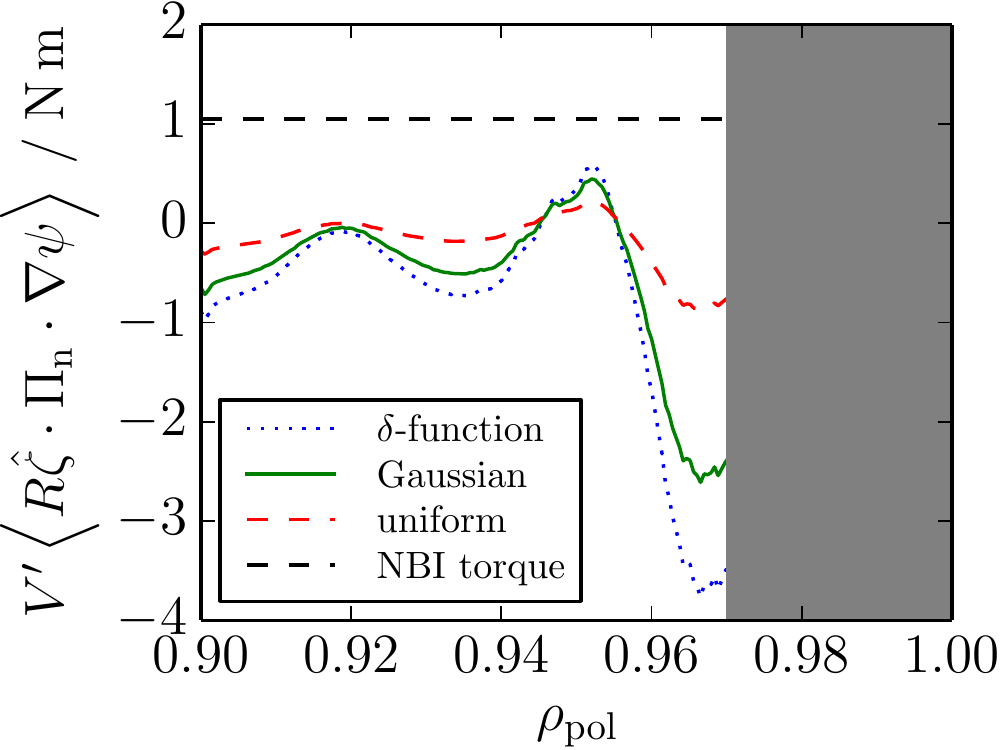}

\caption{Comparison of outward flux of toroidal angular momentum through neutrals
for different poloidal profile shapes: $\delta$-function at outboard
midplane (blue, dotted), Gaussian with width $\pi/5\;\mathrm{rad}$
centred at outboard midplane (green, solid) and uniform (red, dashed).
The total angular momentum flux from the NBI source is indicated as
the horizontal, dashed line. Positive values represent radially outward
flux of co-current angular momentum. Grey shaded area is within a
poloidal gyroradius of the separatrix.\label{fig:mtm-flux}}

\end{figure}

\begin{figure}[tb]
\includegraphics[width=0.5\textwidth]{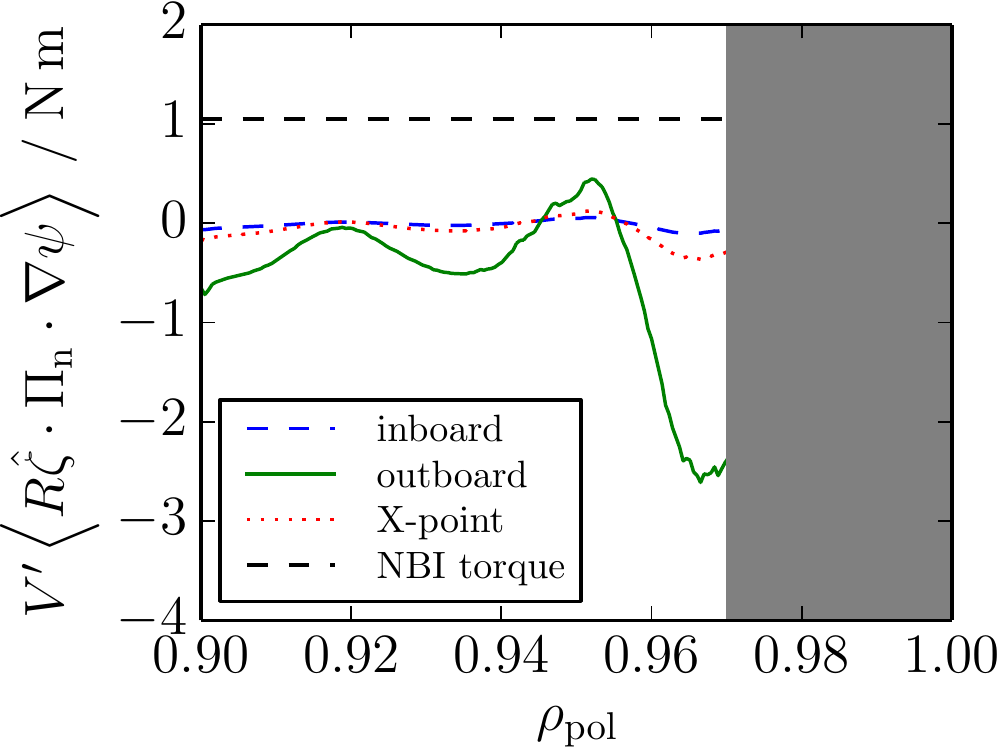}

\caption{Comparison of outward flux of toroidal angular momentum through neutrals
for different poloidal positions: inboard midplane (blue, dashed),
outboard midplane (green, solid) and the nearest poloidal position
to the X-point (red, dotted). Poloidal profile of the neutrals is
Gaussian with width $\pi/5\;\mathrm{rad}$. The total angular momentum
flux from the NBI source is indicated as the horizontal, dashed line.
Positive values represent radially outward flux of co-current angular
momentum. Grey shaded area is within a poloidal gyroradius of the
separatrix.\label{fig:mtm-flux-poloidal-position}}

\end{figure}

To obtain neutral density profiles, we ran KN1D \citep{Labombard2001},
extending the plasma profiles (taking $T_{\mathrm{e}}=\ti$) into
the scrape-off layer as exponentials, with decay lengths of 3~cm
for the density and 0.5~cm for the temperature. KN1D's molecular
pressure input was chosen to give a value of $\nn\approx10^{16}\;\mathrm{m^{-3}}$
at the separatrix, consistent with the low-field side neutral density
modelled in \citep{Viezzer2011}. We use the output of KN1D to set
the flux surface average of the neutral density $\left\langle \nn\right\rangle $,
which we keep fixed while choosing different poloidal profiles in
FIGs.~\ref{fig:mtm-flux} and \ref{fig:mtm-flux-poloidal-position}.

Within a drift-orbit width of the separatrix, the ion distribution
function is likely to depart strongly from the conventional neoclassical
prediction due to orbit loss effects \citep{Shaing1989,Stoltzfus-Dueck12_PoP},
which we do not consider here. We therefore exclude this region in
the presentation of our results, as indicated by the shaded regions
in the figures.

For poloidally uniform neutrals, the outward flux of co-current toroidal
angular momentum passing through a flux surface reaches nearly $-1\;\mathrm{N\, m}$,
as shown in FIG. \ref{fig:mtm-flux}. The discharge had $1.1\;\mathrm{MW}$
of NBI heating with $39.8\;\mathrm{keV}$ particles; the momentum
is deposited at a tangency radius $R_{\mathrm{NBI}}=0.93\;\mathrm{m}$
and so the external momentum input from the NBI is %
\begin{comment}
$E=39.8keV$, $P=1.1MW$. Particle flux $\Gamma=\frac{P}{E}=1.725\times10^{20}s^{-1}$.
Particle momentum is $mv=\sqrt{2mE}=6.53\times10^{-21}kg\, m\, s^{-1}$.
Torque is $\Gamma mvR_{0}=1.05\, N\, m$
\end{comment}
$1.05\;\mathrm{N\, m}$ (indicated as the dashed horizontal line in
FIG.~\ref{fig:mtm-flux}), of a similar magnitude to the momentum
flux carried by the neutrals. This gives the net outward flux of angular
momentum carried by all mechanisms through our simulation domain at
the plasma edge, assuming that the NBI momentum is predominantly deposited
in the core plasma and provides the only source of angular momentum.

The neutrals thus carry a significant angular momentum flux near the
plasma edge and can be expected to play an important role in regulating
the plasma rotation in L-mode.

For comparison one can make a simple estimate of the angular momentum
flux. Assume that the fast neutrals, those coming from a charge exchange
reaction with a hot ion in the confined plasma, escape immediately
from the confined region. Then the total momentum carried out of the
plasma is $\Gamma_{\mathrm{n}}\mi\bar{V}_{\zeta}R_{\mathrm{out}}\approx0.3\;\mathrm{N\, m}$
where $\Gamma_{\mathrm{n}}\approx2\times10^{21}\;\mathrm{s}^{-1}$
is the total inward flux of neutrals (estimated from the flux density
across the separatrix of $10^{20}\;\mathrm{m^{-2}\, s^{-1}}$ simulated
by KN1D), $\bar{V}_{\zeta}\approx20\;\mathrm{km}\,\mathrm{s}^{-1}$
is a typical toroidal rotation velocity and $R_{\mathrm{out}}\approx2\;\mathrm{m}$
is a typical radius in the outboard edge region of the plasma. This
estimate is similar in magnitude to our prediction but has the opposite
sign; this is a consequence of the very different assumptions made.
For this estimate each neutral experiences only one CX interaction
and acts purely as a momentum sink. On the other hand in the short
MFP model we use, each neutral undergoes several CX interactions.
The direction of the flux (inward or outward) is then influenced by
the directions of the gradients in both neutral density and plasma
toroidal rotation, heat flux, etc. because neutrals exchange momentum
between different flux surfaces. In particular there can be an inward
pinch of momentum due to the density gradient of the neutrals; on
a given flux surface there are more neutrals coming from the outside
than the inside, all carrying the momentum picked up from a CX reaction
with an ion. Note that, as we see in FIG.~\ref{fig:profiles}, the
toroidal rotation does not change much across our domain; thus the
momentum carried by each inward or outward going particle will be
similar, so that this effect gives rise to an inward flux of angular
momentum.

Poloidal localization of the neutrals can enhance the angular momentum
flux. In FIG.~\ref{fig:mtm-flux} we compare the momentum flux for
three different assumed poloidal profiles: uniform, Gaussian (with
standard deviation $\pi/5\;\mathrm{rad}$) or $\delta$-function.
The localized profiles are centred at the outboard midplane. For ease
of comparison we have used the same flux-surface averaged neutral
density profile for all of the curves. In reality, the neutral distribution
will be some combination of a roughly uniform piece from wall recycling
and a localized part near the gas fuelling valve (represented by the
Gaussian shape here), so the enhancement above the uniform level due
to outboard localization could be expected to be more moderate than
the limiting case shown. Previous work, both analytical \citep{Fulop02,Helander2003}
and numerical \citep{Omotani2016}, used $\delta$-function localized
neutrals for convenience. We can see from FIG.~\ref{fig:mtm-flux}
that there is not a major difference between this approximation and
the case of neutrals localized as a Gaussian, with both the trends
and order of magnitude being correctly captured, although the sharper
localization of the $\delta$-function does further enhance the momentum
flux.

The neutrals provide more effective transport of toroidal angular
momentum when they are on the outboard side of the tokamak than on
the inboard side or near the X-point, due to the larger major radius
and poloidal magnetic field $B_{\mathrm{p}}$, as can be seen from
(\ref{eq:mtm-flux}), noting that $\left|\nabla\psi\right|=RB_{\mathrm{p}}$.
This is confirmed by FIG. \ref{fig:mtm-flux-poloidal-position} which
compares the momentum flux through Gaussian localized neutrals centred
at the outboard midplane, inboard midplane and adjacent to the X-point,
highlighting the importance of the poloidal location of the neutrals
for the angular momentum flux that they carry.

\begin{comment}
Estimate neutral penetration depth.

At edge, $T_{e}\approx100eV$

$n_{e}\approx9e18m^{-3}$

Guesstimate neutral temperature as $T_{n}\approx100eV$

ionization rate coefficient for this $T_{e}$ is 3.094e-8cm\textasciicircum{}3/s=3.094e-14m\textasciicircum{}3/s

Penetration depth $=\sqrt{\frac{2\times100eV}{3.343583719e-27kg}}/(3.094e-14m^{3}s^{-1}\times9e18m^{-3})\approx$
\end{comment}

\section{Discussion\label{sec:Discussion}}

We have demonstrated here that charge-exchanging neutrals can carry
a significant angular momentum flux in the tokamak edge. The formalism
developed in \citep{Fulop02,Helander2003} and first implemented numerically
in \citep{Omotani2016} has been extended, to enable interpretive
studies evaluating the momentum transport through neutrals for experimental
profiles and equilibria. In this interpretive mode, the radial variation
of the background profiles does not have to be neglected compared
to the radial gradient of the neutral density. We model an L-mode
discharge from AUG to demonstrate this capability and show that a
significant flux of angular momentum, comparable in magnitude to the
total momentum input from NBI, can be carried by the neutrals, motivating
further application to experimental data in future.

The approach taken here allows rapid experimentation with parameters
and profiles and provides both qualitative insight and at least order
of magnitude estimates of the momentum transport due to neutrals.
Despite the limitations of the modelling described here, some conclusions
are clear. The strength of the neutral momentum transport is much
larger for neutrals located on the outboard side of the tokamak than
on the inboard side. This is due both to the smaller moment of force
at smaller major radius and to the smaller physical-space gradients
of flux-function plasma profiles which have poloidally constant gradients
in $\psi$-space, as $\left|\nabla\psi\right|=R\bp$ is smaller on
the inboard side. Likewise near the X-point, where $\bp$ is small,
the influence of the neutrals will be weak.  Thus the influence of
neutrals on the plasma rotation may be minimized by fuelling from
the inboard side or X-point. On the other hand, if it is desired to
drive intrinsic rotation using the neutral angular momentum flux \citep{Fulop02,Helander2003,Omotani2016},
then this will compete more effectively with other momentum transport
channels if the neutrals are located on the outboard side and may
be able to generate substantial radial electric fields in H-mode pedestals.

Intrinsic rotation is particularly important for future tokamaks such
as ITER since the angular momentum source from NBI heating will be
weaker in larger devices. The interesting case is that with steep
temperature profiles in H-mode and although our modelling is restricted
to L-mode plasmas, it is in the region where the edge transport barrier
will form in H-mode that neutrals are most important. However, modelling
steep temperature gradients is extremely challenging since the deviation
of the bulk ion distribution from a Maxwellian distribution is not
small, so that a non-linear collision operator is needed, and the
steep gradients also necessitate radially-global solutions. State
of the art numerical solutions of the neutral kinetic equation couple
only to a drifting-Maxwellian plasma \citep{Reiter2005eirene}. However,
the temperature gradient driven departure of the ions from a local
Maxwellian distribution drives an angular momentum flux in the neutral
distribution function \citep{Catto1994,Helander2003}. Thus when the
external torque is large, so that the rotation-driven momentum flux
dominates, the drifting-Maxwellian description of the bulk plasma
is sufficient, but when there are steep temperature gradients, or
for intrinsic rotation where the external torque vanishes, a kinetic
model for the plasma is needed. In the linear region we have considered,
where standard neoclassical theory and modelling are valid, the transport
is proportional to the gradients, so significantly stronger momentum
transport through neutrals may be expected to arise in steep gradient
regions, such as the H-mode pedestal. Quantitative evaluation, requiring
the development of more sophisticated codes coupling fully kinetic
neutrals to kinetic ions, remains a challenging subject for future
research, which can build on existing progress in neoclassical pedestal
modelling \citep{Catto2013,Landreman14_PERFECT,Battaglia2014,Dorf2016}.

\section*{Acknowledgements}

The authors are grateful to Matt Landreman for advice and help with
the \textsc{perfect} code and to Stuart Henderson for assistance with
the ADAS database. We also thank Prof.~Arne Kallenbach and Dr.~Rachael
McDermott for carefully reading the manuscript. This work was supported
by the Framework grant for Strategic Energy Research (Dnr.~2014-5392)
and the International Career Grant (Dnr.~330-2014-6313) from Vetenskapsr{\aa}det.

{%\setlength{\bibsep}{6pt}
\bibliographystyle{iaea}
\bibliography{references}

}
\end{document}